\documentclass{article}

\usepackage{microtype}
\usepackage{graphicx}
\usepackage{subcaption}
\usepackage{booktabs}
\usepackage{hyperref}

\usepackage[accepted]{icml2026_GenAICreativity}

\usepackage{amsmath}
\usepackage{amssymb}
\usepackage{mathtools}
\usepackage{amsthm}
\usepackage[capitalize,noabbrev]{cleveref}
\usepackage{enumitem}
\usepackage{xcolor}
\usepackage{textcomp}
\usepackage[textsize=tiny]{todonotes}
\usepackage{tcolorbox}

\icmltitlerunning{Creativity from Friction: Human--AI Interaction for Exploratory Structural Design}

\begin{document}

\twocolumn[
  \icmltitle{Creativity from Friction: \\Human--AI Interaction for Exploratory Structural Design}

  \begin{icmlauthorlist}
    \icmlauthor{Ricardo Maia Avelino}{eth1,eth2}
    \icmlauthor{Rita Sevastjanova}{eth2}
    \icmlauthor{Tom Van Mele}{eth1}
    \icmlauthor{Philippe Block}{eth1}
    \icmlauthor{Mennatallah El-Assady}{eth2}
  \end{icmlauthorlist}

  \icmlaffiliation{eth1}{Block Research Group, Department of Architecture, ETH Zurich, Switzerland}
  \icmlaffiliation{eth2}{IVIA Lab, Department of Computer Science, ETH Zurich, Switzerland}

  \icmlcorrespondingauthor{Ricardo Maia Avelino}{mricardo@ethz.ch}

  \icmlkeywords{Human-AI collaboration, structural design, Human-AI co-creation design space exploration, constrained creativity, human sketches, architecture}

  \vskip 0.3in
]

\printAffiliationsAndNotice{}

\begin{abstract}
AI agents that generate final answers based on user input often do not meet the needs of creative fields. Fields such as structural design and architecture need interactive systems that help users externalise and develop ideas, explore alternatives, and refine partial solutions. The final product of such designs needs to comply with many constraints concerning, e.g., spatial configuration, mechanical behaviour, material quantities, and costs. These constraints create friction in the design process, which can stimulate novel and creative solutions. In this paper, we discuss the misalignment between current generative AI goals to remove friction and provide final solutions and the needs of creators, such as structural designers, who develop ideas through iterative work. We present the design dimensions of systems allowing for \emph{constrained human--AI co-creation} that rely on vision-language models making structural exploration conversational, multimodal, and responsive to evolving human intent in ways that follow and augment the discipline's creative process. Through a pilot design interface based on these principles and a study with experts in the field, this paper shows how structural designers perceive interactive AI systems and how such systems can support design space exploration by reducing repetitive modelling friction while preserving reflective design friction.

\end{abstract}

\section{Introduction}
\label{sec:introduction}

Generative AI is often evaluated for its ability to produce plausible final outputs such as an image, a text, a plan, or a model from user input. However, creative design work rarely progresses as a one-shot process to the final product. Designers externalise partial ideas, compare alternatives, and use intermediate models to think, collaborate, and create. This view resonates with Sch{\"o}n's model for professionals of creative disciplines that engage in ``reflection-in-action'' adapting their creation process to ever evolving constraints and ideas \citep{schon1983reflective}. Thus, generative AI in creative domains must shift from workflow automation to \emph{exploration augmentation}~\cite{rezwana2023designing, wang2025exploring}.

Structural design makes this need particularly visible. It is a creative discipline, but not an unconstrained one. Structural engineers must balance spatial constraints while reasoning about loads, supports, spans, constructability, economy, safety, among others. In doing so, engineers seek creative solutions that emerge from the friction of managing all these constraints. \citet{teng2004fostering} define creativity in structural design as the production of a design that is novel and satisfies functional constraints, and highlight the importance of divergent thinking to draw analogies from solutions to previous problems and explore different alternatives.

Structural exploration has been the focus of recent research in which algorithms that explore different structural forms are preferred rather than single-objective optimisations on predetermined typologies \citep{mueller2014computational, DANHAIVE2021103664, lee2016automatic,saldanaochoa2021beyond}. Generative AI brings new opportunities to expand this exploration by allowing multimodal interactions and advanced reasoning \cite{zhang2025exploring, suh2024luminate}. This raises a key question: \textit{how can we formally define such interactions so that these systems effectively support and enhance the design process?}

We answer this question by introducing the concept of \textit{constrained co-creation} for structural design: interactive human–AI workflows in which creative exploration is shaped by the ongoing negotiation of domain constraints. The design process supported by these systems becomes an interactive activity in which human and AI agents jointly explore and refine evolving design alternatives. Unlike final-answer automation, constrained co-creation helps designers externalise, inspect, revise, and evaluate alternatives while maintaining human agency \cite{holter2024deconstructing} and allowing domain-grounded interactive workflows \citep{song2024human}. Such systems can remove friction from repetitive actions while maintaining friction that arises from the constraints and steers the design process.

This paper contributes to human--AI interaction in the field of structural design by (1) defining productive friction and its importance for interactive structural design, (2) identifying four design dimensions for future human--AI systems, (3) presenting a prototype of an AI--supported design workflow, and (4) reporting a preliminary study to investigate how experts interact with this prototype in a proposed design scenario.

\section{Structural Design Exploration and Human--AI Co-Creation}
\label{sec:framing}

\subsection{Structural Design as Creative Constraint~Negotiation}

\begin{figure}[t]
    \centering
    \includegraphics[width=0.92\linewidth]{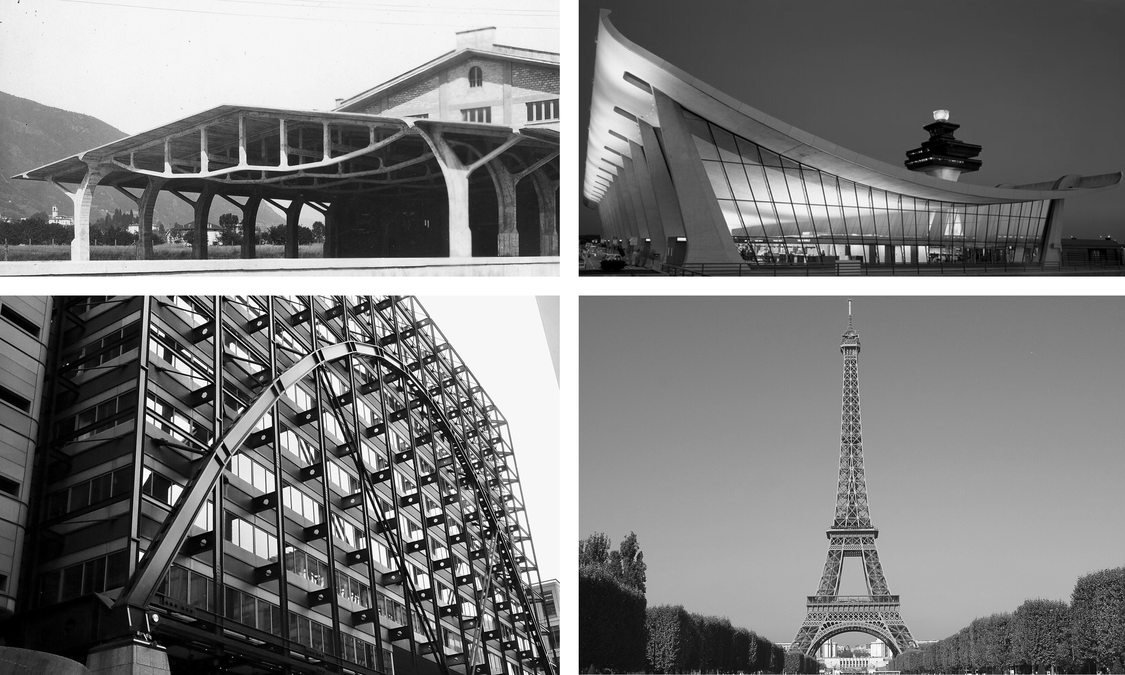}
    \caption{Examples of Billington's structural art: Chiasso Shed, Switzerland, by Robert Maillart (Photo: ETH-Bibliothek Zürich); Dulles Airport, United States, by Eero Saarinen (Photo: Carol Highsmith); Broadgate Exchange House, United Kingdom, by Skidmore, Owings \& Merrill (Photo: Wikimedia Commons); and the Eiffel Tower, France, by Gustave Eiffel (Photo: Wikimedia Commons).}
    \label{fig:creative-structures}
    \vspace{-2em}
\end{figure}

Structural design is the discipline that connects architecture and physics and is concerned with ensuring the mechanical performance of building systems that safely and efficiently resist applied loads while satisfying architectural, spatial, material, environmental, and construction requirements. Unlike many creative domains in which an artefact may be judged primarily by aesthetic or expressive qualities, a structural design is inseparable from how it works. Billington defines structural ``art" as a creative synthesis of the three E's of efficiency, economy, and elegance \citep{billington1983tower}. Figure~\ref{fig:creative-structures} illustrates constructed examples of Billington's structural art where the structures are not conceived by an unconstrained creative act but are rather the result of the negotiation of form and force under multiple and multidisciplinary constraints \citep{allen2010form}. Among them, the Chiasso Shed truss geometry follows its internal forces allowing for minimal material usage; the Dulles Airport roof follows the geometry of a catenary resulting in an efficient purely-tensile geometry; the Broadgate Exchange House embeds a load-bearing arch to transfer forces away from the centre, clearing the train tracks running below the building; and the Eiffel Tower was shaped to resist the effects of lateral wind loads. These novel projects, achieved through creative and constrained design space exploration, remain landmark works today.

\subsection{AI-Supported Design Space Exploration}

Computational structural design has long treated early-stage design as the exploration of a space of possibilities rather than the production of a single optimal answer. Shape grammars are often used to encode design transformations and generate families of spatial configurations \citep{stiny1980introduction}. In structural design, grammars are used in \citet{shea1999novel, lee2016automatic}, allowing the generation of large design spaces navigated through stochastic algorithms. \citet{vanmele2012geometry, maiaavelinoInteractiveImplementationAlgebraic2021} develop graphic-statics-based algorithms to explore the interactive design of 2D structures. \citet{ohlbrock2016cem} propose a combinatorial approach that allows exploration and form-finding of spatial configurations in equilibrium for 3D systems. The latter approach is further developed in \citet{saldanaochoa2021beyond, pastranaConstrainedFormfindingTension2023, guo2026text2structure3d}, allowing for projection-based design exploration, introducing auto-differentiation algorithms for equilibrium, and using graph transformers to generate structures from text descriptions.

Recent work on generative AI for building structures extends this trajectory by surveying how data representations, generative algorithms, evaluation methods, and optimisation loops are being used for structural design automation \citep{liao2024}. However, much of this work still frames AI primarily as a generator or optimiser. Language and vision-language agents, in contrast, offer new opportunities to support a conversational and multimodal form of design space exploration. Instead of presenting designers with precomputed or parametrised alternatives, such agents can allow users to steer the generation, transformation, and evaluation of structural models through text and multimodal input. Indeed, the design intent in architecture and structural engineering is rarely communicated through text alone. Designers use sketches, annotations, and partial models to express ambiguous spatial ideas. Recent work on sketch-based 3D generation highlights both the promise and the difficulty of interpreting sketches in engineering \citep{tono2025deep,li2025llm4cad,he2025}. Sketches are useful because they are fast and expressive and sometimes encode tacit knowledge, but they are also incomplete, ambiguous, and difficult for current models to interpret reliably. The challenge is therefore to build AI systems that can work with incomplete multimodal human expressions while keeping the evolving artefact grounded in structural logic.

\subsection{Design Dimensions for Human--AI Structural Design Interfaces}

We derive four design dimensions for human–AI interfaces to support exploratory structural design following the guidelines proposed in \citet{stahle2025design} for agentic interfaces in visual analytics. These dimensions specialise previous co-creative and mixed-initiative frameworks \citep{rezwana2023designing, wang2025exploring, holter2024deconstructing} to the field of structural design. We propose that these interfaces should support: (1) model grounding in structural design knowledge, (2) human- and AI-readable data structures, (3) state awareness and interaction history, and (4) multimodal expression of design intent.

\paragraph{Model grounding in structural design knowledge.}
A reliable system must be grounded in discipline-specific logic \cite{song2024human}. For structural design, this includes basic notions of structural load paths, knowledge of typical spans, loads, and cross-sections, and understanding of member hierarchy and boundary condition definition. Without such a grounding, an AI system may generate artefacts that are visually convincing but structurally incorrect, which might, in turn, cause domain experts to lose confidence in the system.
\paragraph{Human- and AI-readable data structures.}~The design artefact should be represented in a form that is accessible to both humans and AI systems \cite{cao2025compositional,feng2026cocoa}. For the human, the model should be visual, inspectable, and editable. For AI, the data that underlie the artefacts need to be accessible, revealing their internal data structures and interconnections. This shared representation enables editing for both humans and AI. 
\paragraph{State awareness and interaction history.}
A co-creative AI system must enable access, inspection, and modification of the current state of an evolving design model \cite{feng2026cocoa}. As the artefacts develop, constraints must be re-evaluated. An edit may remove a support, disconnect an element, or change the structural load path. The system should therefore maintain awareness of the current state, previous edits, persistent constraints, and user goals to effectively assist in the design of safety-critical engineering components and to allow tracing the responsibility 
over executed modifications.

\paragraph{Multimodal expression of design intent.}
Structural and architectural intent is often expressed in many ``languages'' \cite{shi2026talksketch,baudoux2025multimodal}. Designers communicate through sketches, drawings, annotations, and verbal or text descriptions. Communication is also often incomplete and ambiguous. Human–AI tools should therefore accommodate these distinct ``languages'' and their intrinsic ambiguity.

Together, these dimensions can guide the role of AI in exploratory structural design. The next section proposes a workflow that embeds these design dimensions.

\section{A Co-creation Structural Design Workflow}
\label{sec:workflow}

\begin{figure*}[t]
    \centering
    \includegraphics[width=0.92\textwidth]{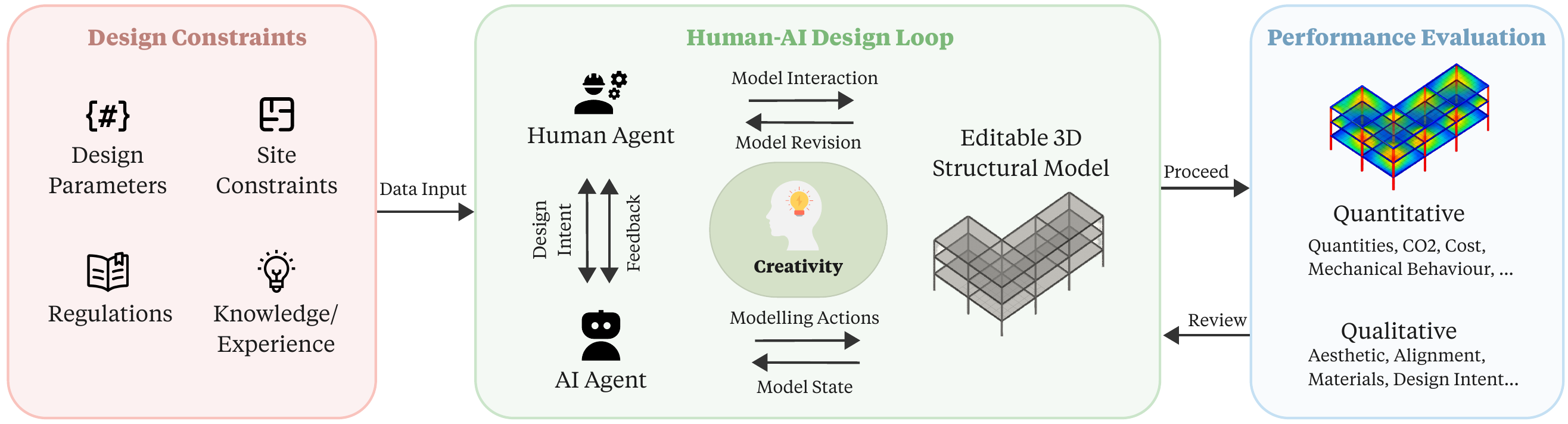}
    \caption{Proposed human--AI workflow for exploratory structural design. The interactions between human, AI, and the model support creative exploration under a constrained design space.}
    \label{fig:workflow}
    \vspace{-1.5em}
\end{figure*}

Based on the dimensions presented above, this work proposes a co-creative workflow for exploratory structural design allowing for human$\rightleftarrows$model$\rightleftarrows$AI interaction and anchored on pre-defined and intrinsic discipline constraints (Figure~\ref{fig:workflow}). Explicit constraints, such as design parameters, site conditions, and regulations, are defined at the beginning of the process. These are complemented by intrinsic discipline constraints based on engineering validity or knowledge. Some of this knowledge should be embedded in the model's grounding, while some remains part of the designer's tacit expertise and professional judgement about what is likely to perform well. Motivated by these constraints, the human and AI agents \textit{co-create} in a design loop where the design intent is co-developed between them. Through multimodal inputs, both agents can edit and interact with the model. Due to interactivity and state awareness, both agents can perceive the model changes and review the current state. The agents share bi-directional feedback throughout the design process.

We argue that this cycle can support an augmented design exploration, which might lead to more creative ways to satisfy the constraints. The goal is not to automate the production of a final structural form but to follow the design process, reducing unproductive modelling friction while preserving the \textit{productive friction} through which designers inspect, revise, and negotiate constraints. By editing the structural model, receiving feedback from the AI agent, the human agent acquires important insights that influence the design process. The generated designs can then be evaluated for quantitative or qualitative performance and the results might reveal problems that then act as additional constraints requiring new creative solutions.

To understand the effectiveness and potential of this design workflow, we conducted a preliminary study on a pilot structural design interface that enables bi-directional multimodal interactive structural design using Google Gemini 3.1 Flash-Lite \citep{Gemini31Flashlite} as its underlying reasoning model. The study is described in the following section.

\section{Preliminary Design Study}
\label{sec:study}

We conduct a preliminary study to examine how expert users interact with an AI-supported structural design interface in a constrained design scenario. This study offers qualitative insight into how designers use, trust, revise, and respond to an AI agent during early-stage structural design and how they manage friction in the design process.

\subsection{Design Task}

\begin{figure}[!b]
    \centering
    \includegraphics[width=0.75\linewidth]{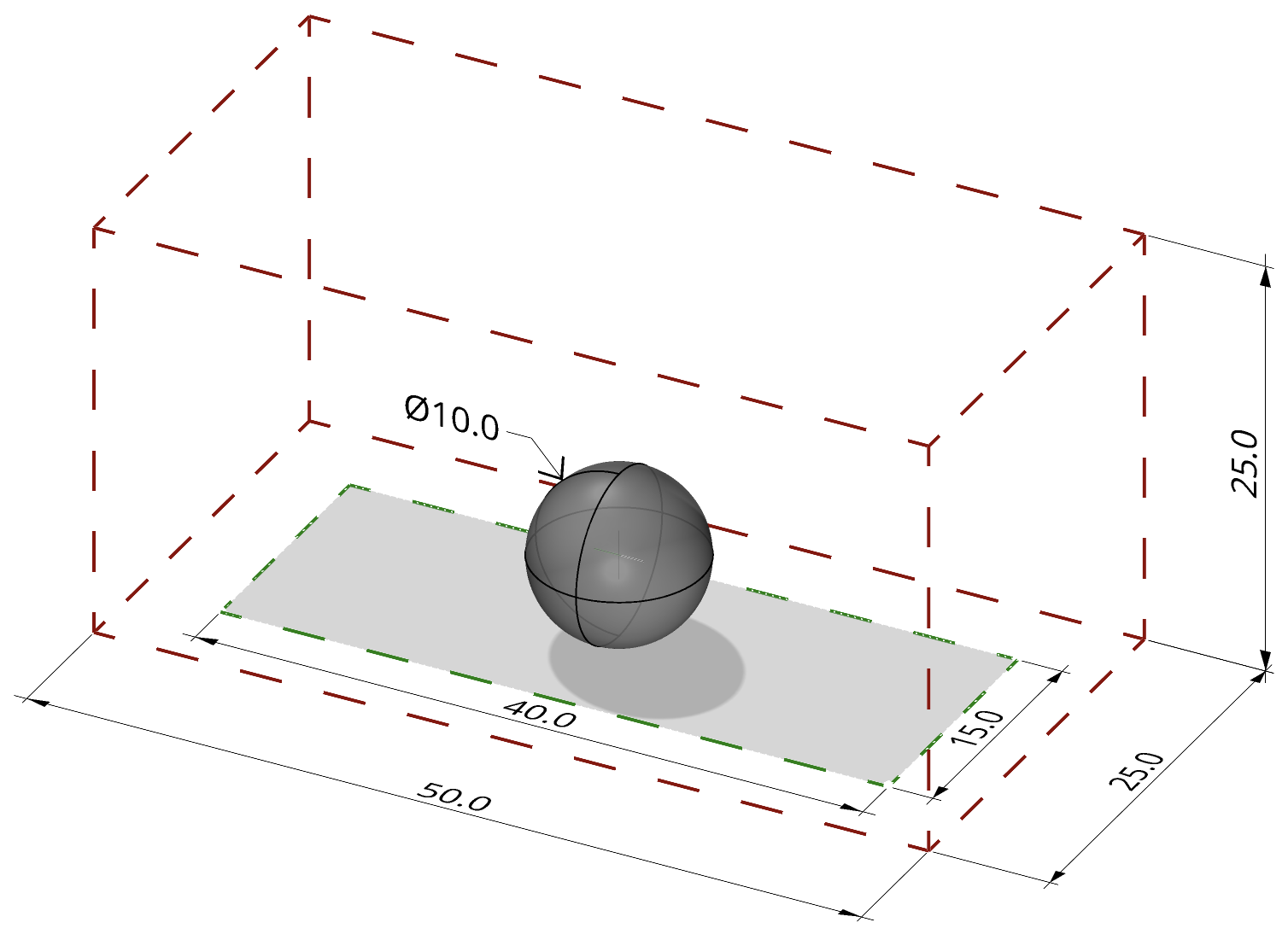}
    \caption{Boundaries of the user study. The building must stay within the red boundary, avoid the spherical object, and be supported in the green zone. Measurements are indicated in metres.}
    \label{fig:study-task}
\end{figure}

The participants are given the design task shown in Figure~\ref{fig:study-task}. The task proposes to design a multi-storey building within a maximum limit of $50.0\times 25.0\times 25.0\,\mathrm{m}$ with supports only allowed in a reduced footprint of $40.0\times 15.0\,\mathrm{m}$. In addition, the building must provide an open space in the middle to allow a public plaza such that the building cannot cross a virtual sphere with diameter of $10.0\,\mathrm{m}$ positioned in the centre of the support footprint. Designers are instructed to develop creative but structurally feasible solutions with at least 5 storeys and $2000\,\mathrm{m}^2$ floor area. 

\subsection{Participants and Procedure}

We recruit $n=3$ participants with experience in the fields of architecture, structural engineering, and parametric modelling. Each session starts with a consent form and then proceeds in six stages guided by a questionnaire (Appendix~\ref{app:interview}). A test facilitator observes and assists participants, especially when they struggle or feel uncertain about using the AI interface, following the pair-analytics design approach \cite{arias2011pair}. The stages are defined as:

\begin{enumerate}[leftmargin=*,nosep]
    \item \textbf{Demographics:} Participants fill out a survey of their professional and educational experience and their knowledge of 3D modelling and AI (Appendix~\ref{app:demographics}).
    \item \textbf{Task presentation:} Participants receive the design problem and printed pages of the orthographic views of the building and are asked to develop their ideas by sketching their design intent (Appendix~\ref{app:full-sketches}).
    \item \textbf{Introduction to the tool:} Participants receive a 5 minute introduction of the tool and how they can create and edit artefacts using drafting commands, AI conversation, or a sketch-on-the-screen tool.
    \item \textbf{AI co-creation:} Participants interact with the system through direct 3D modelling, text prompts to AI, and sketch-based AI inputs, iteratively refining the design. Participants' screen activities are recorded (Figure~\ref{fig:interaction}).
    \item \textbf{Final result:} The participants review the final artefact and approve it (Figure~\ref{fig:final-models}).
    \item \textbf{Reflection interview:} Participants discuss their experience and how the interaction with AI modified their design output and where it reduced or created friction (Section~\ref{sec:discussion}).
\end{enumerate}

\subsection{User Interaction}

\begin{figure*}[t]
    \centering
    \includegraphics[width=\textwidth]{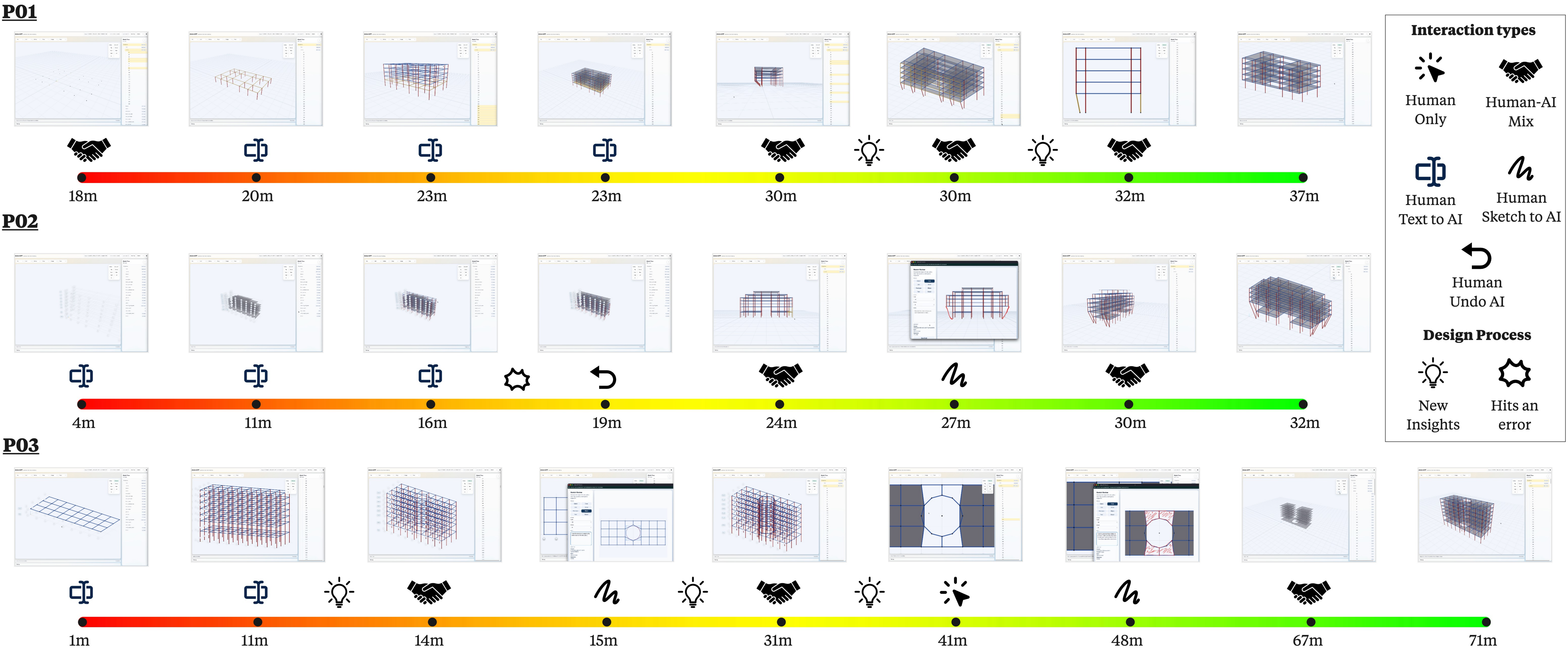}
    \caption{Evolving design timeline for all participants (P01, P02 and P03) using the 3D interface. Interactions are characterised and key moments in the design process are noted.}
    \label{fig:interaction}
    \vspace{-1.5em}
\end{figure*}

This section describes the interaction of each participant with the AI interface. Figure~\ref{fig:interaction} summarises the evolution of the design sessions, showing key intermediate model states, interaction types, and approximate timestamps for each participant. The interactions are characterised as follows:

\begin{itemize}[nosep]
    \item Human-Only: modifications primarily executed by the human alone using the 3D drafting commands.
    \item Human-AI-Mix: modifications are shared between humans and AI, e.g. ``move selected joints by 4m.''
    \item Human-Text-to-AI: high level prompts written by participants requiring modifications in the structure, e.g. ``populate beams and columns on a grid with sizes [...].''
    \item Human-Sketch-to-AI: human performs a sketch on the screen which is sent together with a prompt-instruction asking for AI modifications.
    \item Human-Undo-AI: AI makes unexpected mistakes and the user intervenes and undo.
\end{itemize}

The interaction for each participant is described below.

\paragraph{P01.}
\textit{Sketch Phase (18 min):} Based on the constraints received, P01 developed their sketch mainly at the front elevation, ensuring that the area and floor constraints were fulfilled. The concept included inclined columns on the ground floor and an enlarged atrium to accommodate the open space constraint. 
\textit{3D Phase (37 min):} P01 combined manual geometry modifications with AI assisted modelling operations. The participant began defining ground-level joints at the intended column locations, populated a beam grid, and then asked AI to replicate them to the upper floors. The participant asked AI to remove the ``largest'' slab to meet the open-space constraint. After reviewing the AI output, the inclined columns were rejected as the participant judged that these members were too long. The participant then moved the south face inward to reduce the inclined column lengths but later noticed that these would end up too close to the typical columns for practical engineering construction. This insight changed the design direction with fewer columns and a $2.5\,\mathrm{m}$ overhang at north and south sides. The participant approved the final result after checking the area constraints with AI assistance.

\paragraph{P02.}
\textit{Sketch Phase (15 min):} The participant's concept presented inclined columns along the shortest building direction; after reviewing their first sketch, the participant decided to add a staggering on the top floors to make the design more unique. Front and side sketches were made. 
\textit{3D Phase (32 min):} P02 started the 3D modelling with a clearer predefined concept, and AI was used primarily to accelerate the construction of the structural scheme. The participant created a grid and described the staggering using the grid tag names. The slabs were created correctly on the first attempt, but AI struggled to extend the same logic consistently to beams and columns. The participant then restarted with a revised prompt that successfully generated most of the required elements. The participant then used sketches to indicate the desired inclined columns at the building edges to the AI and made further small edits using text prompts. The final model was then approved and finalised after performing an AI check on the area constraint. 

\paragraph{P03.}
\textit{Sketch Phase (16 min):} P03 developed the sketch in plan using a regular $5.0\,\text{m} \times 5.0\,\text{m}$ grid with terraces on the longest side of the structure. The open-space constraint was envisaged as a minimal cutout in the building regular grid at low level columns.
\textit{3D Phase (71 min):} P03's 3D interaction took longer than for other participants. P03 started collaborating with the AI system to generate the regular building. However, P03 changed their direction when planning the cutout for the sphere. Using the sketch tool, P03 found that a top-to-bottom cylindrical cutout would be more appealing than the original planned minimal cutout. Since the regular building grid was already set, the interaction of P03 with the model became time consuming, requiring passing specific instructions to the AI system to rearrange specific elements of the building. P03 used the sketch feature to indicate the generation of the irregular slab shapes created by the cutout. P03 was satisfied with the final outcome, but noted that additional time would have allowed to also include the terraces as in the original idea. 

In summary, participants interacted with AI on different levels. P01 modified its initial design intent but this change was not suggested by the AI autonomously, but rather by the participant’s visual inspection of AI changes. P02 used AI mainly as an accelerator of its design intent, retaining responsibility for correcting and refining the spatial logic. P03's interaction shows the promise and limitation of multimodal co-creation since sketching helped the participant express an irregular spatial idea, but the AI struggled to operationalise it precisely. 

\begin{figure}[t]
    \centering
    \includegraphics[width=0.99\linewidth]{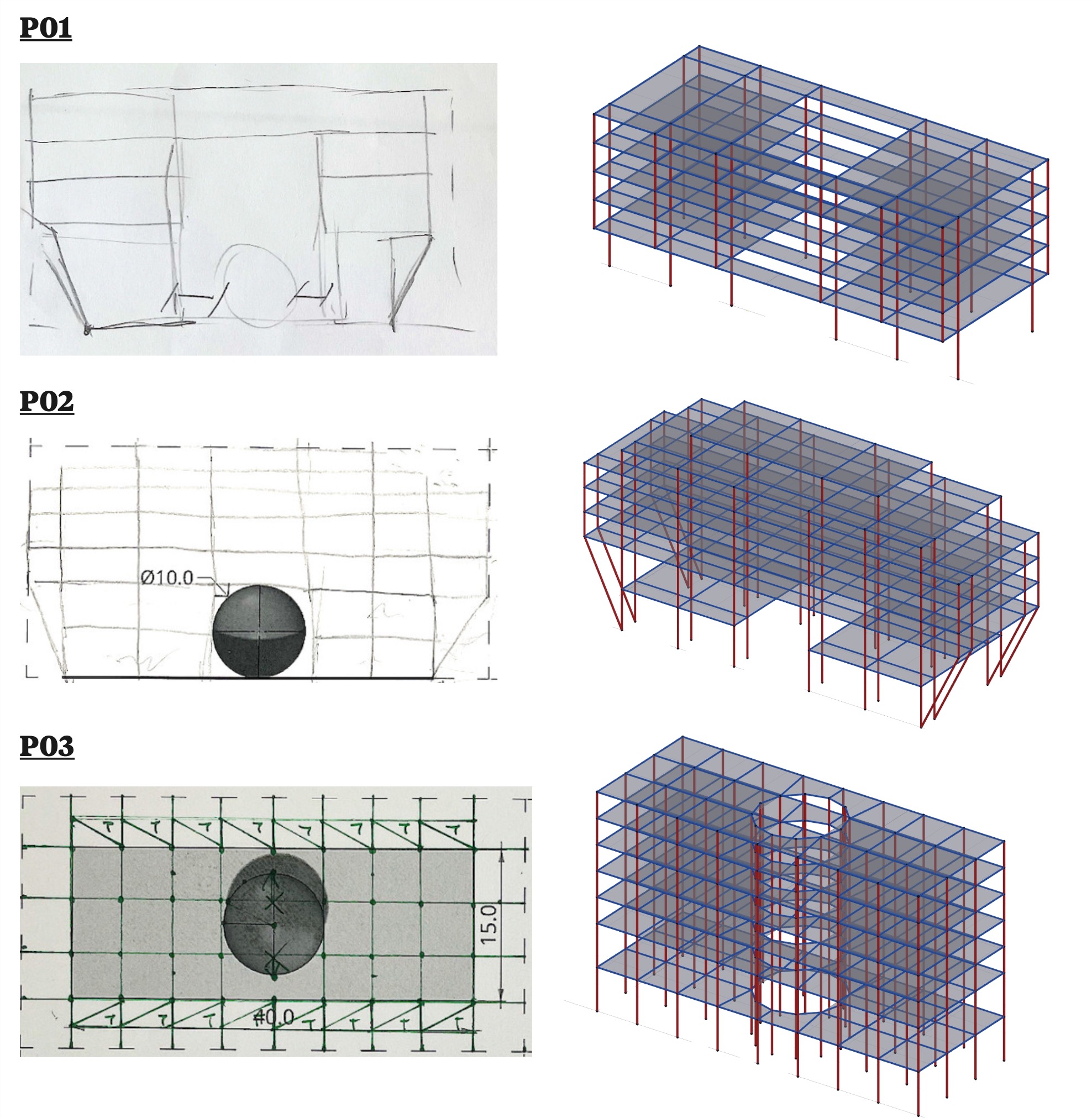}
    \caption{Initial sketch and final models co-developed using the AI tool for participants P01, P02 and P03.}
    \label{fig:final-models}
    \vspace{-1.5em}
\end{figure}

The initial sketch and the final 3D model for all participants are presented in Figure~\ref{fig:final-models}. Following their 3D modelling session, participants discussed the post-task interview questions which are presented in the next section.

\section{Discussion: AI for Creative Structural Design}
\label{sec:discussion}

\subsection{Creativity as the journey, not the destination}

The central value of AI in creative domains may not be the production of a final answer, but the expansion and navigation of possible alternatives. In the present study, we observed that all participants have modified their original design idea either in the manual sketch phase or during the 3D modelling session. P03 reported that the AI-supported process helped develop the idea of a circular internal court from the ground floor to the roof, which had not been considered before. P01 changed their design idea twice, achieving a design that successfully met the constraints. P02 also changed their design idea in the sketching phase, erasing the top bays of the building to create a more interesting staggering effect. This interaction shows that professionals in creative domains find new solutions as they develop their original designs. An AI system that acts as a co-creative assistant can, therefore, enrich the design journey and spark new creative insights from structural designers.

\subsection{Productive and Unproductive Friction}

Not all friction should be removed from creative work. Some friction is productive since it forces designers to confront constraints, compare alternatives, and make judgements. Other friction, however, is unproductive, including repetitive modelling actions, interface overhead, and low-level translation between intent and geometry.

In our study, all participants identified the reduction of repetitive modelling as an area where AI was helpful. P01 emphasised that AI quickly created repetitive elements. P02 noted that the AI made it possible to quickly generate a structure from a grid and remove unwanted bays through simple prompts, avoiding manual removal that would otherwise have taken longer. P03 similarly highlighted the rapid generation of the main structural grid and floor slabs. This indicates that the current prototype was effective in executing explicit, bounded, and repetitive modelling operations.

Similarly, the interface has maintained the productive friction. By interacting with the interface, P01 noticed two design flaws, first regarding the length of the inclined columns and then identifying supports too close to each other. As a result, P01 has changed the design to remove these elements altogether, resulting in a simpler, elegant, and still feasible solution. True co-creation systems should move towards identifying and removing unproductive friction while preserving or highlighting reflective friction that will support design judgement. 


\subsection{Multimodality as a Co-Creative Anchor}  

Multimodal input, such as sketches and drawings, can guide exploration without over-specifying it. The ambiguity of sketches lets participants express spatial intent and design directions more naturally than natural language alone. In our study, all participants indicated that they use sketches as one of the main modals to start a new design. P02 added that starts thinking about projects in 2D and then goes to 3D using CAD software and that the pair of 3D and sketches helps to fully understand the problem. 

However, while P01 also responded that using sketches felt natural, P01 did not use the feature during the session. P01 added that many repetitive operations, such as replicating multiple elements at a given distance, can be more easily communicated using natural language. On the other hand, P02 noted that it was sometimes hard to choose a vocabulary to communicate with AI. P02 noted that ``it was hard to put my ideas into words for the prompt, similarly to how it is hard to explain a design idea to someone without drawing.'' 

These experiences confirm that multimodality is key for a co-creative system. Some tasks are more easily described in prompts, while for others a sketch might as well be worth a thousand words.

\subsection{Automation, Collaboration, or Co-creation?}

Users were asked whether they would categorise their interaction as automation, collaboration, or co-creation, and why. The reflection of the participants revealed different expectations about the agency in human--AI structural design workflows. P01 categorised the current interaction as automation, describing co-creation as an AI system capable of issuing early warnings about feasibility or problematic design decisions. P02 described the interaction as collaboration because both the human and AI can modify the model. For P02, co-creation means an AI system that would ``suggest things automatically'' or highlight when the design was drifting from the original constraints. P03 also described the current experience as a collaboration and appreciated being in control of the design process and ``figuring out new ideas'' independently as the project progressed. These responses support the argument that co-creation in structural design does not necessarily mean greater AI autonomy. Instead, it requires a careful distribution of agency which should adapt to the user's preference as discussed in \citep{holter2024deconstructing}.

\subsection{Limitations}

This paper presents a position and a preliminary study rather than an exhaustive evaluation of structural validity. The participant sample was small, even if it allowed one to identify interaction patterns and design dimensions. The interface is in continuous development, so not all drawing functions comparable to CAD software are currently supported. In particular, users have reported missing functions of click-and-drag and snap that could facilitate the action of human and AI agents. Furthermore, in the present study, changes in the design direction were initiated mainly by designers themselves rather than AI. The current study also does not establish that AI use alone caused more creative outcomes. Instead, it identifies interaction patterns between designers and structural artefacts that can guide how AI-powered interfaces might more effectively support co-creation. Future work could study this process subjected to increased AI autonomy, where it could proactively suggest actions or signal potential design redirections.

\section{Conclusion}
\label{sec:conclusion}

We argue that creative AI systems for structural design, as well as for other creative disciplines, should not be designed as final-answer generators, but rather as interactive partners for exploring the design space. Observing participants interact with a prototype AI interface, this paper shows how design intent and user objective evolve together with the artefact. The final artefact is then shaped by the constraints of the design problem and by the intrinsic feasibility constraints of structural design. Future AI-assisted structural design systems should then follow this process, assisting by removing unproductive friction coming from modelling repetition while preserving productive friction that support designers in achieving their final outcome. 

We argue that for structural design, constraints have a dual effect: they reduce the design space while also creating opportunities to stimulate creativity. We propose design dimensions for future human--AI tools for structural design including \textit{model grounding} in discipline-specific knowledge, \textit{human- and AI-readable data structures} allowing both agents to modify the model, \textit{state awareness and interaction history} enabling real-time inspection and traceability, and \textit{multimodal expression of design intent} for a more natural and flexible workflow. The preliminary study described in this paper has reinforced the importance of these design dimensions, and future work will focus on further interface development, systematic evaluation of AI systems for structural design, exploration of different levels of AI autonomy and improving domain specific model grounding. We believe that future work on these unique challenges of adapting generative AI for creative tasks can unlock new possibilities towards a more creative and interactive co-design process for engineers.

\section*{Acknowledgments}
\label{sec:acknowledgement}

This work was supported by the Design++ Postdoctoral Fellowship awarded to the first author in collaboration with Halter AG.

\bibliography{references}

@book{allen2010form,
  title     = {{Form and Forces: Designing Efficient, Expressive Structures}},
  author    = {Allen, Edward and Zalewski, Wac{\l}aw},
  year      = {2010},
  publisher = {John Wiley \& Sons},
  address   = {Hoboken, NJ}
}

@inproceedings{arias2011pair,
  title     = {{Pair Analytics: Capturing Reasoning Processes in Collaborative Visual Analytics}},
  author    = {Arias-Hernandez, Richard and Kaastra, Linda T. and Green, Tera M. and Fisher, Brian},
  booktitle = {{2011 44th Hawaii International Conference on System Sciences}},
  pages     = {1--10},
  year      = {2011},
  publisher = {IEEE},
  doi       = {10.1109/HICSS.2011.339}
}

@article{baudoux2025multimodal,
  title     = {{Multimodal Generative {AI} for Conceptual Design: Enabling Text-Based and Sketch-Based Human--AI Conversations}},
  author    = {Baudoux, Ga{\"e}lle and Guo, Chenjun and Goucher-Lambert, Kosa},
  journal   = {{Proceedings of the Design Society}},
  volume    = {5},
  pages     = {2501--2510},
  year      = {2025},
  doi       = {10.1017/pds.2025.10264}
}

@book{billington1983tower,
  title     = {{The Tower and the Bridge: The New Art of Structural Engineering}},
  author    = {Billington, David P.},
  year      = {1983},
  publisher = {Princeton University Press},
  address   = {Princeton, NJ}
}

@inproceedings{cao2025compositional,
  title     = {{Compositional Structures as Substrates for Human--AI Co-Creation Environment: A Design Approach and a Case Study}},
  author    = {Cao, Yining and Huang, Yiyi and Truong, Anh and Shin, Hijung Valentina and Xia, Haijun},
  booktitle = {{Proceedings of the 2025 CHI Conference on Human Factors in Computing Systems}},
  series    = {{CHI '25}},
  year      = {2025},
  articleno = {188},
  numpages  = {25},
  pages     = {1--25},
  publisher = {Association for Computing Machinery},
  address   = {New York, NY, USA},
  doi       = {10.1145/3706598.3713401}
}

@article{guo2026text2structure3d,
      title={{Text2Structure3D: Graph-Based Generative Modeling of Equilibrium Structures with Diffusion Transformers}}, 
      author={Lazlo Bleker and Zifeng Guo and Kaleb Smith and Kam-Ming Mark Tam and Karla Saldaña Ochoa and Pierluigi D'Acunto},
      journal={arXiv preprint arXiv:2601.12870},
      year={2026},
      eprint={2601.12870},
      archivePrefix={arXiv},
      primaryClass={cs.CE},
      url={https://arxiv.org/abs/2601.12870}, 
}

@article{DANHAIVE2021103664,
  title   = {{Design Subspace Learning: Structural Design Space Exploration Using Performance-Conditioned Generative Modeling}},
  author  = {Danhaive, Renaud and Mueller, Caitlin T.},
  journal = {{Automation in Construction}},
  volume  = {127},
  pages   = {103664},
  year    = {2021},
  doi     = {10.1016/j.autcon.2021.103664}
}

@inproceedings{feng2026cocoa,
  title     = {{Cocoa: Co-Planning and Co-Execution with {AI} Agents}},
  author    = {Feng, K. J. Kevin and Pu, Kevin and Latzke, Matt and August, Tal and Siangliulue, Pao and Bragg, Jonathan and Weld, Daniel S. and Zhang, Amy X. and Chang, Joseph Chee},
  booktitle = {{Proceedings of the 2026 CHI Conference on Human Factors in Computing Systems}},
  series    = {{CHI '26}},
  year      = {2026},
  articleno = {16},
  numpages  = {23},
  pages     = {1--23},
  publisher = {Association for Computing Machinery},
  address   = {New York, NY, USA},
  doi       = {10.1145/3772318.3791673}
}

@misc{Gemini31Flashlite,
  title        = {{Gemini 3.1 Flash-Lite: Model Card}},
  author       = {{Google DeepMind}},
  year         = {2026},
  howpublished = {\url{https://deepmind.google/models/model-cards/gemini-3-1-flash-lite/}},
  note         = {Accessed: 2026-05-15}
}

@article{he2025,
  title   = {{Generative {AIBIM}: An Automatic and Intelligent Structural Design Pipeline Integrating {BIM} and Generative {AI}}},
  author  = {He, Zhili and Wang, Yu-Hsing and Zhang, Jian},
  journal = {{Information Fusion}},
  volume  = {114},
  pages   = {102654},
  year    = {2025},
  doi     = {10.1016/j.inffus.2024.102654}
}

@article{holter2024deconstructing,
  title   = {{Deconstructing Human--AI Collaboration: Agency, Interaction, and Adaptation}},
  author  = {Holter, Steffen and El-Assady, Mennatallah},
  journal = {{Computer Graphics Forum}},
  volume  = {43},
  number  = {3},
  pages   = {e15107},
  year    = {2024},
  doi     = {10.1111/cgf.15107}
}

@article{lee2016automatic,
  title   = {{Automatic Generation of Diverse Equilibrium Structures through Shape Grammars and Graphic Statics}},
  author  = {Lee, Juney and Mueller, Caitlin and Fivet, Corentin},
  journal = {{International Journal of Space Structures}},
  volume  = {31},
  number  = {2--4},
  pages   = {147--164},
  year    = {2016},
  doi     = {10.1177/0266351116660798}
}

@article{li2025llm4cad,
  title   = {{{LLM4CAD}: Multimodal Large Language Models for Three-Dimensional Computer-Aided Design Generation}},
  author  = {Li, Xingang and Sun, Yuewan and Sha, Zhenghui},
  journal = {{Journal of Computing and Information Science in Engineering}},
  volume  = {25},
  number  = {2},
  pages   = {021005},
  year    = {2024},
  doi     = {10.1115/1.4067085}
}

@article{liao2024,
  title   = {{Generative {AI} Design for Building Structures}},
  author  = {Liao, Wenjie and Lu, Xinzheng and Fei, Yifan and Gu, Yi and Huang, Yuli},
  journal = {{Automation in Construction}},
  volume  = {157},
  pages   = {105187},
  year    = {2024},
  doi     = {10.1016/j.autcon.2023.105187}
}

@inproceedings{maiaavelinoInteractiveImplementationAlgebraic2021,
  title     = {{An Interactive Implementation of Algebraic Graphic Statics for Geometry-Based Teaching and Design of Structures}},
  author    = {Maia Avelino, Ricardo and Lee, Juney and Van Mele, Tom and Block, Philippe},
  booktitle = {{International fib Symposium: Conceptual Design of Structures 2021}},
  pages     = {447--454},
  year      = {2021},
  address   = {Attisholz-Areal, Switzerland},
  doi       = {10.35789/fib.PROC.0055.2021.CDSymp.P054}
}

@phdthesis{mueller2014computational,
  title  = {{Computational Exploration of the Structural Design Space}},
  author = {Mueller, Caitlin T.},
  school = {{Massachusetts Institute of Technology}},
  year   = {2014}
}

@article{ohlbrock2016cem,
  title   = {{Combinatorial Equilibrium Modeling}},
  author  = {Ohlbrock, Patrick Ole and Schwartz, Joseph},
  journal = {{International Journal of Space Structures}},
  volume  = {31},
  number  = {2--4},
  pages   = {177--189},
  year    = {2016},
  doi     = {10.1177/0266351116660799}
}

@article{pastranaConstrainedFormfindingTension2023,
  title   = {{Constrained Form-Finding of Tension--Compression Structures Using Automatic Differentiation}},
  author  = {Pastrana, Rafael and Ohlbrock, Patrick Ole and Oberbichler, Thomas and D'Acunto, Pierluigi and Parascho, Stefana},
  journal = {{Computer-Aided Design}},
  volume  = {155},
  pages   = {103435},
  year    = {2023},
  doi     = {10.1016/j.cad.2022.103435}
}

@article{rezwana2023designing,
  title     = {{Designing Creative {AI} Partners with {COFI}: A Framework for Modeling Interaction in Human--AI Co-Creative Systems}},
  author    = {Rezwana, Jeba and Maher, Mary Lou},
  journal   = {{ACM Transactions on Computer-Human Interaction}},
  volume    = {30},
  number    = {5},
  articleno = {67},
  numpages  = {28},
  pages     = {1--28},
  year      = {2023},
  publisher = {Association for Computing Machinery},
  address   = {New York, NY, USA},
  doi       = {10.1145/3519026}
}

@article{saldanaochoa2021beyond,
  title   = {{Beyond Typologies, Beyond Optimization: Exploring Novel Structural Forms at the Interface of Human and Machine Intelligence}},
  author  = {Saldana Ochoa, Karla and Ohlbrock, Patrick Ole and D'Acunto, Pierluigi and Moosavi, Vahid},
  journal = {{International Journal of Architectural Computing}},
  volume  = {19},
  number  = {3},
  pages   = {277--299},
  year    = {2021},
  doi     = {10.1177/1478077120943062}
}

@book{schon1983reflective,
  title     = {{The Reflective Practitioner: How Professionals Think in Action}},
  author    = {Sch{\"o}n, Donald A.},
  year      = {1983},
  publisher = {Basic Books},
  address   = {New York}
}

@article{shea1999novel,
  title   = {{The Design of Novel Roof Trusses with Shape Annealing: Assessing the Ability of a Computational Method in Aiding Structural Designers with Varying Design Intent}},
  author  = {Shea, Kristina and Cagan, Jonathan},
  journal = {{Design Studies}},
  volume  = {20},
  number  = {1},
  pages   = {3--23},
  year    = {1999},
  doi     = {10.1016/S0142-694X(98)00019-2}
}

@inproceedings{shi2026talksketch,
  title     = {{TalkSketch: Multimodal Generative {AI} for Real-Time Sketch Ideation with Speech}},
  author    = {Shi, Weiyan and Upadhyay, Sunaya and Quek, Geraldine and Choo, Kenny Tsu Wei},
  editor    = {Woodward, Kieran and Falcon-Caro, Alicia and Ramchurn, Richard and Benford, Steve},
  booktitle = {{Creative {AI} for Live Interactive Performances}},
  pages     = {83--97},
  year      = {2026},
  publisher = {Springer Nature Switzerland},
  address   = {Cham},
  doi       = {10.1007/978-3-032-16893-1_6}
}

@article{song2024human,
  title     = {{Human--AI Collaboration by Design}},
  author    = {Song, Binyang and Zhu, Qihao and Luo, Jianxi},
  journal   = {{Proceedings of the Design Society}},
  volume    = {4},
  pages     = {2247--2256},
  year      = {2024},
  publisher = {Cambridge University Press},
  doi       = {10.1017/pds.2024.227}
}

@article{stahle2025design,
  title   = {{A Design Space for Intelligent Agents in Mixed-Initiative Visual Analytics}},
  author  = {St{\"a}hle, Tobias and Jansen op de Haar, Matthijs and Boyer, Sophia and Sevastjanova, Rita and Narechania, Arpit and El-Assady, Mennatallah},
  journal = {{arXiv preprint arXiv:2512.23372}},
  year    = {2025},
  url     = {https://arxiv.org/abs/2512.23372}
}

@article{stiny1980introduction,
  title   = {{Introduction to Shape and Shape Grammars}},
  author  = {Stiny, George},
  journal = {{Environment and Planning B: Planning and Design}},
  volume  = {7},
  number  = {3},
  pages   = {343--351},
  year    = {1980},
  doi     = {10.1068/b070343}
}

@inproceedings{suh2024luminate,
  title     = {{Luminate: Structured Generation and Exploration of Design Space with Large Language Models for Human--AI Co-Creation}},
  author    = {Suh, Sangho and Chen, Meng and Min, Bryan and Li, Toby Jia-Jun and Xia, Haijun},
  booktitle = {{Proceedings of the 2024 CHI Conference on Human Factors in Computing Systems}},
  series    = {{CHI '24}},
  year      = {2024},
  articleno = {644},
  numpages  = {26},
  pages = {1--26},
  publisher = {Association for Computing Machinery},
  address   = {New York, NY, USA},
  doi       = {10.1145/3613904.3642400}
}

@article{teng2004fostering,
  title   = {{Fostering Creativity in Students in the Teaching of Structural Analysis}},
  author  = {Teng, J. G. and Song, Chang Yong and Yuan, Xing Fei},
  journal = {{International Journal of Engineering Education}},
  volume  = {20},
  number  = {1},
  pages   = {96--102},
  year    = {2004}
}

@article{tono2025deep,
  title   = {{Deep Sketch-Based {3D} Modeling: A Survey}},
  author  = {Tono, Alberto and Wu, Jiajun and Wetzstein, Gordon and Armeni, Iro and Subramonyam, Hariharan and Landay, James and Fischer, Martin},
  journal = {{Computer Graphics Forum}},
  pages   = {e70302},
  year    = {2025},
  doi     = {10.1111/cgf.70302}
}

@article{vanmele2012geometry,
  title   = {{Geometry-Based Understanding of Structures}},
  author  = {Van Mele, Tom and Lachauer, Lorenz and Rippmann, Matthias and Block, Philippe},
  journal = {{Journal of the International Association for Shell and Spatial Structures}},
  volume  = {53},
  number  = {174},
  pages   = {285--295},
  year    = {2012}
}

@article{wang2025exploring,
  title   = {{Exploring Creativity in Human--AI Co-Creation: A Comparative Study across Design Experience}},
  author  = {Wang, Nan and Kim, Hyunsuk and Peng, Junfeng and Wang, Jiayi},
  journal = {{Frontiers in Computer Science}},
  volume  = {7},
  pages   = {1672735},
  year    = {2025},
  doi     = {10.3389/fcomp.2025.1672735}
}

@article{zhang2025exploring,
  title     = {{Exploring Collaboration Patterns and Strategies in Human--AI Co-Creation through the Lens of Agency: A Scoping Review of the Top-Tier {HCI} Literature}},
  author    = {Zhang, Shuning and Wang, Hui and Yi, Xin},
  journal   = {{Proceedings of the ACM on Human-Computer Interaction}},
  volume    = {9},
  number    = {7},
  articleno = {CSCW413},
  numpages  = {43},
  pages     = {1--43},
  year      = {2025},
  publisher = {Association for Computing Machinery},
  address   = {New York, NY, USA},
  doi       = {10.1145/3757594}
}
\bibliographystyle{icml2026}

\newpage
\appendix
\onecolumn

\clearpage
\section{Appendix: User Study Questionnaire}
\label{app:interview}

Participants completed a questionnaire before and after the design session. The questionnaire collected background information, prior experience, familiarity with relevant tools, and reflections on the AI-assisted design workflow. The study evaluated the prototype interaction rather than the participants' design ability.

\subsection{Consent}

Before participating, participants were asked to confirm consent for participation, analysis of interaction data, use of anonymised quotes, use of anonymised screenshots, sketches or design artefacts, and screen recording during the session.

\subsection{Participant Background}

Participants were first asked to describe their background. They select among multiple fields of expertise from the following options:

\begin{itemize}[leftmargin=*,nosep]
    \item Architecture
    \item Architecture-engineering
    \item Structural engineering
    \item Civil engineering
    \item Computational design
    \item Design research
    \item Engineering research
    \item Architecture student
    \item Engineering student
    \item Other
\end{itemize}

The participants were then asked to report their years of experience in the following areas:

\begin{itemize}[leftmargin=*,nosep]
    \item Structural design practice
    \item Architectural design practice
    \item Computational design or parametric modelling
    \item Engineering studies
    \item Architecture studies
\end{itemize}

The response options for these questions were:

\begin{quote}
None; less than 1 year; 1--3 years; 3--5 years; 5--10 years; more than 10 years.
\end{quote}

\subsection{Familiarity With Tools and Workflows}

Participants rated their familiarity with relevant tools and workflows using a five-point scale:

\begin{quote}
1 = never used; 2 = tried once or twice; 3 = occasional user; 4 = frequent user; 5 = expert or professional user.
\end{quote}

The following categories were included:

\begin{itemize}[leftmargin=*,nosep]
    \item 3D modelling tools, such as Rhino, SketchUp, Blender, or Revit
    \item Parametric modelling tools, such as Grasshopper or Dynamo
    \item Structural analysis or FEM tools, such as RFEM, SAP2000, ETABS, Karamba, or Abaqus
    \item AI conversational tools, such as ChatGPT, Claude, or Gemini
    \item AI tools for design, CAD, modelling, or coding
    \item Sketching as part of the participant's design process
\end{itemize}

\subsection{Pre-Task Questions}

Before starting the design task, participants answered questions about their usual early-stage design process and their expectations for AI assistance.

\begin{enumerate}[leftmargin=*,nosep]
    \item How often do you explore multiple structural alternatives before converging on a concept?
    \item When starting an early-stage structural design problem, which inputs do you usually rely on?
    \item What do you expect an AI assistant to be useful for in early-stage structural design?
\end{enumerate}

For the first question, participants used a five-point scale from ``never'' to ``always.'' For the second question, participants could select multiple options including:

\begin{itemize}[leftmargin=*,nosep]
    \item Sketches
    \item Plans or drawings
    \item Physical models
    \item Previous 3D models
    \item Precedent images
    \item Parametric tools
    \item Other
\end{itemize}

\subsection{Post-Task Reflection Questions}

After completing the AI-assisted design session, participants reflected on how the interaction affected their design process. The questions focused on design evolution, exploration, friction, agency, constraint reasoning, communication modes, and perceived level of collaboration.

\begin{enumerate}[leftmargin=*,nosep]
    \item Did the interaction with the 3D interface modify your initial design idea? If so, how?
    \item Did the AI help you explore alternatives you would not have considered alone? If so, which alternatives?
    \item Did the AI anticipate next steps in a similar way as you would have done alone? If so, when?
    \item Where did the AI remove repetitive modelling friction?
    \item Where did the AI introduce new friction?
    \item Did you feel in control of the design direction?
    \item Did the AI help on your reasoning about constraints? If so, how?
    \item Which mode did you find most useful and natural to communicate with the AI?
    \item Would you describe the interaction as automation, collaboration, or co-creation? Why?
    \item Do you have any additional comments?
\end{enumerate}

Post-task questions used a five-point scale.

\begin{quote}
1 = not at all / strongly disagree; 5 = very much / strongly agree.
\end{quote}

For the communication, participants selected one of the following options:

\begin{itemize}[leftmargin=*,nosep]
    \item Natural-language only
    \item Natural-language + sketch
    \item Sketch only
    \item Direct modification of the 3D interface
    \item Other
\end{itemize}

\section{Appendix: Participant demographics responses.}
\label{app:demographics}

\renewcommand{\thefigure}{B.\arabic{figure}} 
\setcounter{figure}{0}                      

\begin{figure}[h]
    \centering
    \includegraphics[width=0.99\linewidth]{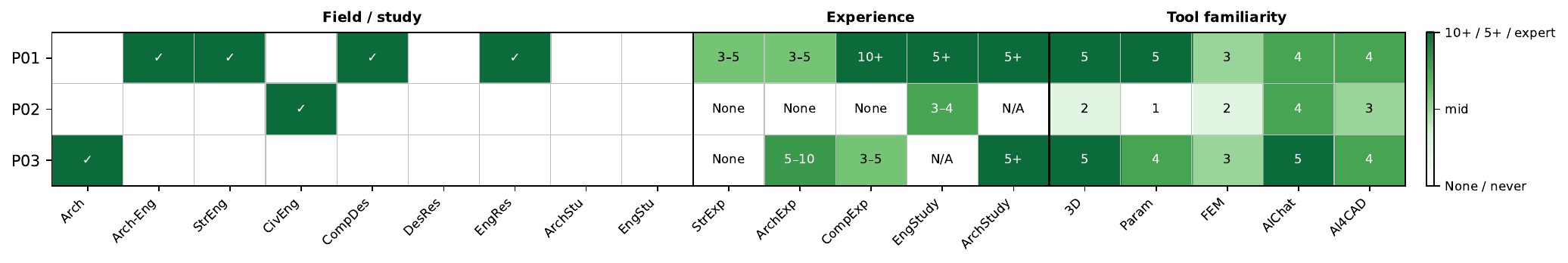}
    \caption{Summary of the demographic results and experience collected from the participants.}
    \label{fig:placeholder}
\end{figure}

\clearpage
\section{Appendix: Sketches created by the participants}
\label{app:full-sketches}

\renewcommand{\thefigure}{C.\arabic{figure}} 
\setcounter{figure}{0}                      

The sketches provided by the participants are shown in this appendix.

\begin{figure}[h!]
    \centering

    \makebox[\textwidth][c]{%
        \begin{tabular}{@{}c@{\hspace{0.06\textwidth}}c@{}}
            \includegraphics[
                height=0.41\textheight,
                keepaspectratio
            ]{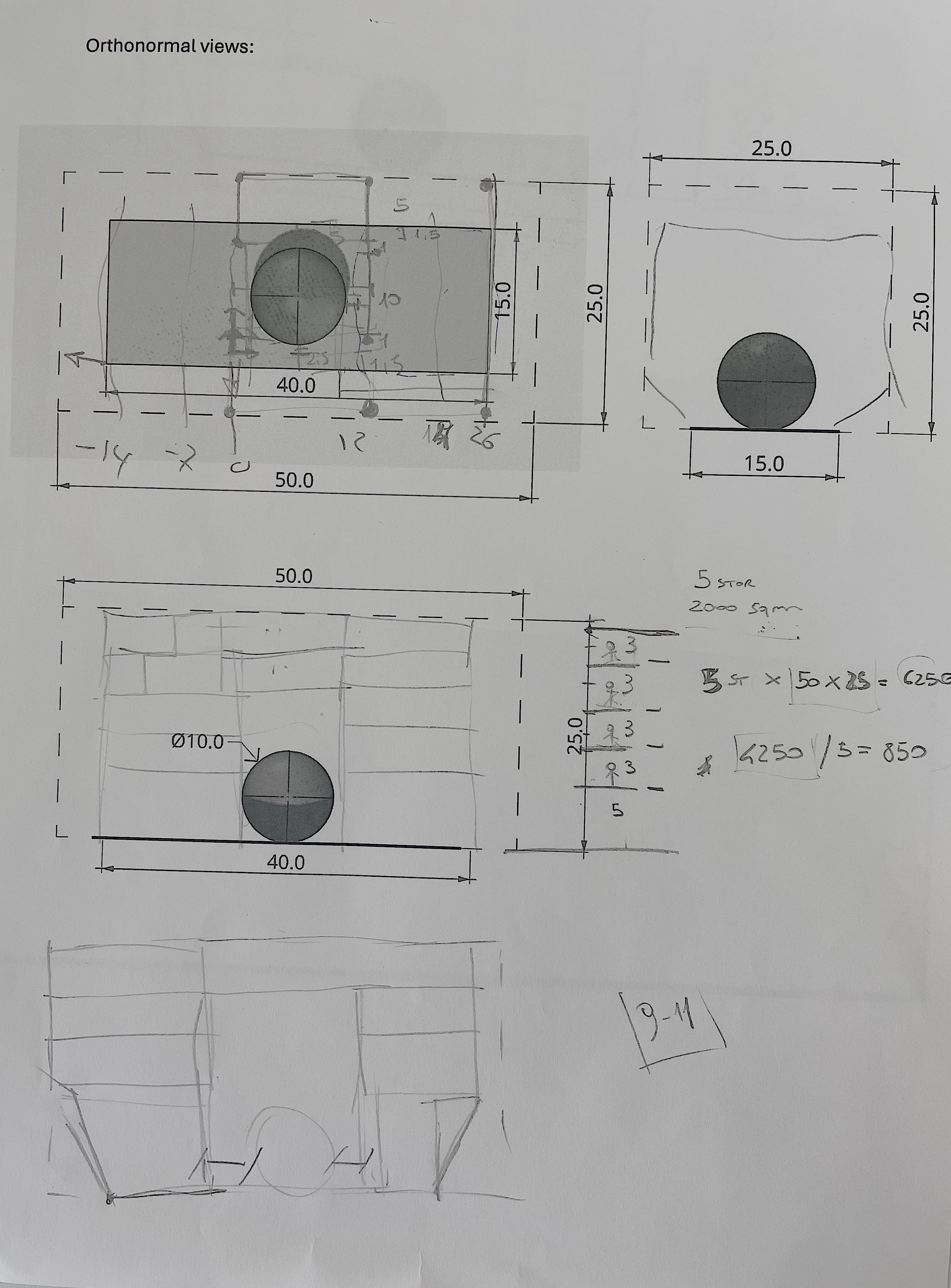}
            &
            \includegraphics[
                height=0.41\textheight,
                keepaspectratio
            ]{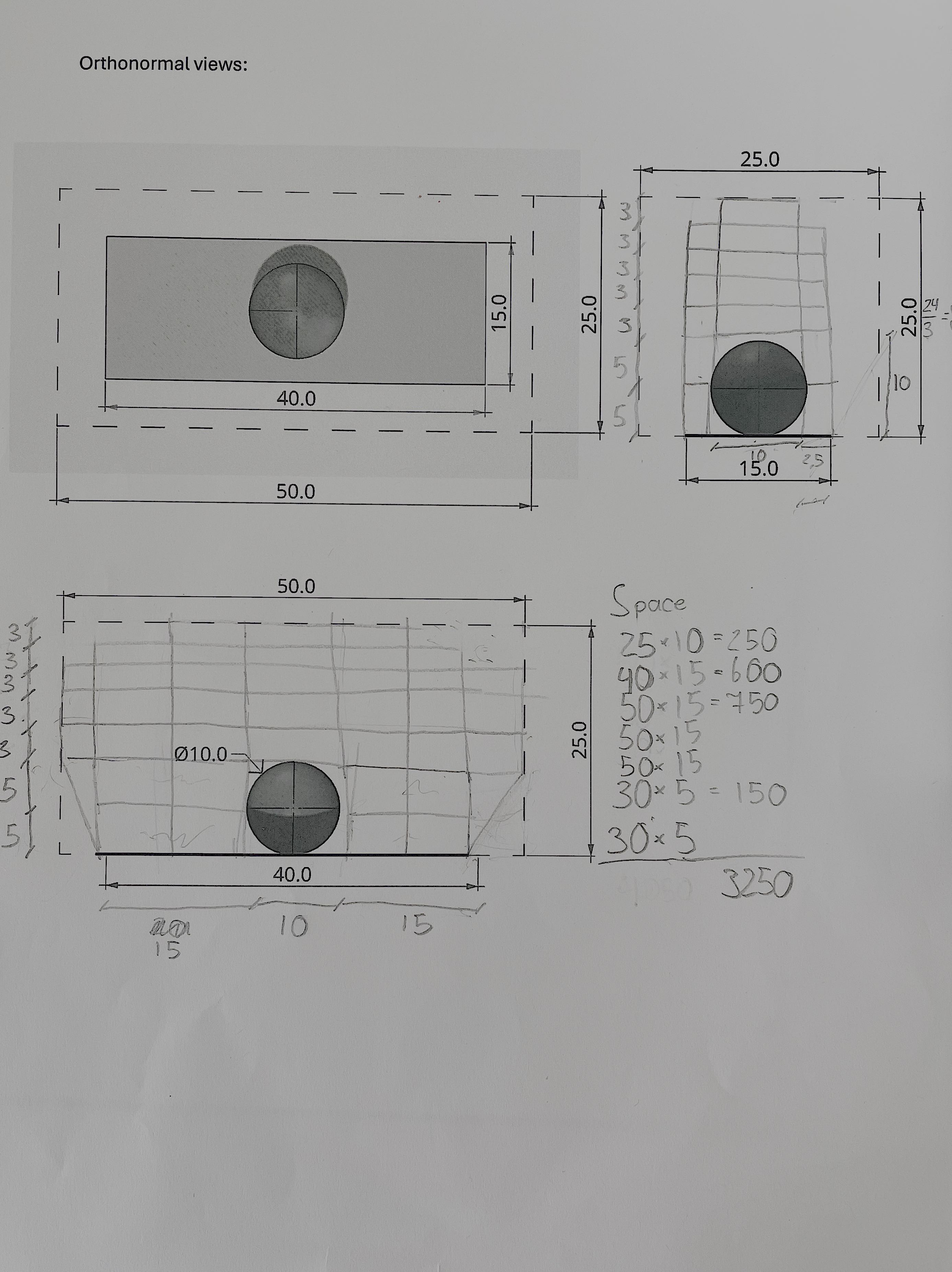}
            \\
            \small (a) Participant P01
            &
            \small (b) Participant P02
        \end{tabular}
    }

    \vspace{0.6em}

    \includegraphics[
        height=0.41\textheight,
        keepaspectratio
    ]{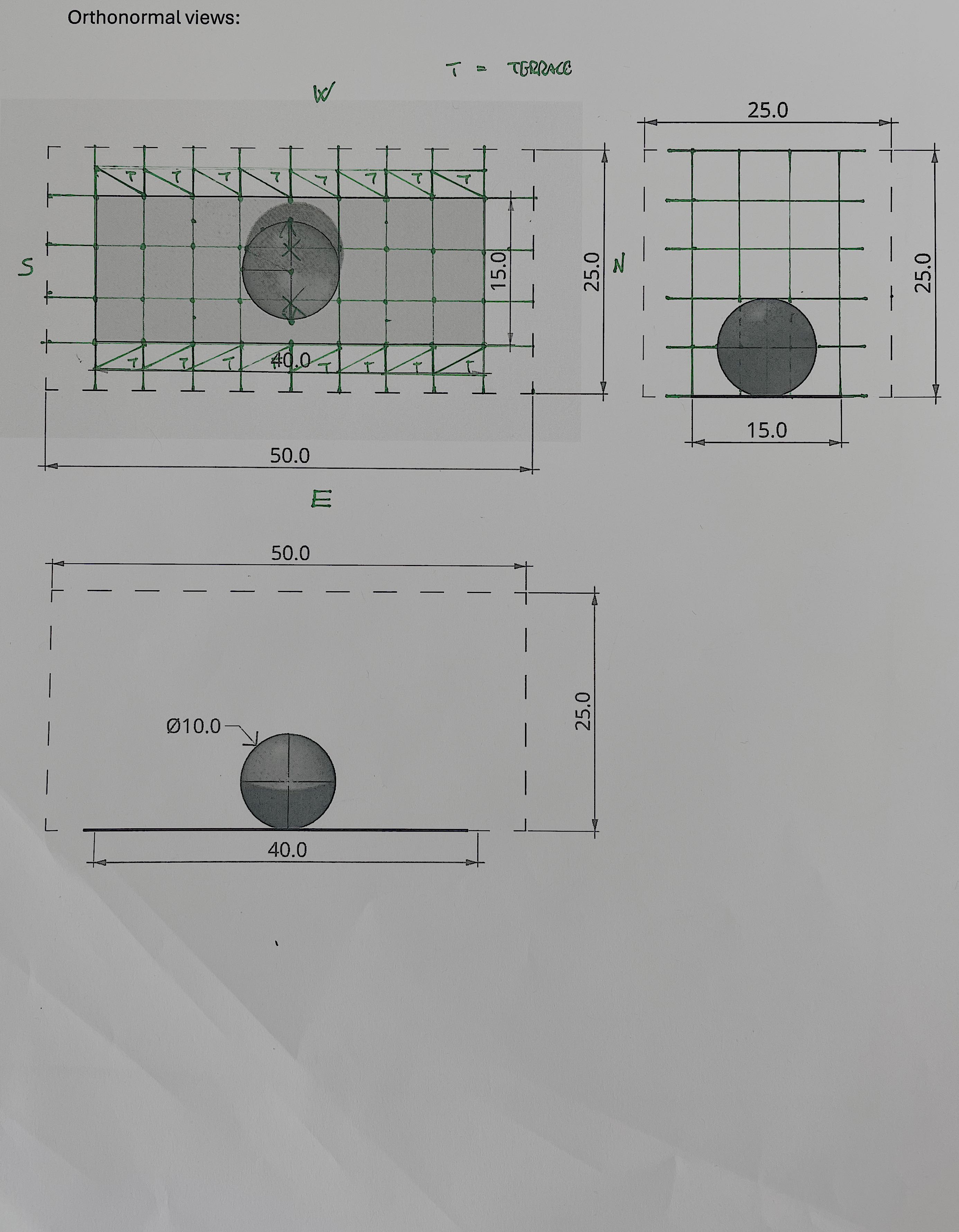}

    \vspace{0.2em}

    {\small (c) Participant P03}

    \caption{Sketches made by the participants during the sketch session.}
    \label{fig:participant-sketches}
\end{figure}

\end{document}